\documentclass[12pt,american,english]{amsart}
\usepackage[T1]{fontenc}
\usepackage{a4wide}
\usepackage{graphicx}
\usepackage{verbatim}

 \theoremstyle{plain}    
 \newtheorem{thm}{Theorem}[section]
 \numberwithin{equation}{section} %% Comment out for sequentially-numbered
 \numberwithin{figure}{section} %% Comment out for sequentially-numbered
 \theoremstyle{plain}    
 \newtheorem{pro}[thm]{Proposition} %%Delete [thm] to re-start numbering
 \newtheorem{lem}[thm]{Lemma} %%Delete [thm] to re-start numbering
\usepackage{babel}

\begin{document}
\def\eps{\epsilon}
\def\tr{{\rm tr}}
\def\i{{\rm i}}
\def\e{{\rm e}}
\def\defi{\stackrel{\rm def}{=}}
\def\t2{{\mathbb T}^2}
\def\k{\mathbf{k}}
\def\N{\mathbb N}
\def\Z{\mathbb Z}
\def\R{\mathbb R}
\def\C{\mathbb C}
\def\O{\mathcal{O}}
\def\S{\mathcal{S}}
\def\hh{\mathcal{H}_h}
\def\AW{\ensuremath{\scriptstyle AW}}
\def\hto0{\xrightarrow{h\to 0}}
\def\htoo{\stackrel{h\to 0}{\longrightarrow}}
\def\rto0{\xrightarrow{r\to 0}}
\providecommand{\abs}[1]{\lvert#1\rvert}
\providecommand{\norm}[1]{\lVert#1\rVert}

\title{On the maximal scarring for quantum cat map eigenstates}
\author{Frédéric Faure \and Stéphane Nonnenmacher}
\address{Laboratoire de Physique et Mod\'elisation des Milieux Condens\'es (LPM2C),
BP 166, 38042 Grenoble C\'edex 9, France ({\tt frederic.faure@ujf-grenoble.fr}, 
{\tt http://lpm2c.grenoble.cnrs.fr/faure})}
\address{Service de Physique Th\'eorique,
CEA/DSM/PhT Unit\'e de recherche associ\'ee au CNRS CEA/Saclay 91191
Gif-sur-Yvette c\'edex, France ({\tt nonnen@spht.saclay.cea.fr})}
\address{Mathematical Sciences Research Institute, 1000 Centennial Drive, Berkeley, 
CA 94720-5070, USA ({\tt nonnenma@msri.org})}
\date{\today}

\begin{abstract}
We consider the quantized hyperbolic automorphisms on the $2$-dimensional torus (or generalized
quantum cat maps), and study the localization properties of their eigenstates in phase
space, in the
semiclassical limit. We prove that if the semiclassical measure corresponding to a sequence
of normalized eigenstates has a pure point component (phenomenon of ``strong scarring''), 
then the weight of this component cannot be larger than the weight of 
the Lebesgue component, and therefore admits the sharp upper bound $1/2$.
\end{abstract}
\maketitle

%%%%%%%%%%%%%%%%%%%%%%%%%%%%%%%%%%%%%%%%%%%%%%%%%%%%%%%%%%%%%%%%%%%%%%%%%%%%%%%
%%%%%%%%%%%%%%%%%%%%%%%%%%%%%%%%%%%%%%%%%%%%%%%%%%%%%%%%%%%%%%%%%%%%%%%%%%%%%%%

\section{Introduction}
We are interested in the semiclassical properties of quantum maps on the two-dimensional torus $\t2$,
that is the unitary transformations $\hat M_h$ which quantize a symplectic map $M$ on $\t2$
($h=2\pi\hbar$ is Planck's constant). 
More precisely, we focus of maps $M$ having a chaotic dynamics (that is, at least ergodic w.r.to
the Lebesgue measure), and 
investigate the phase space properties of the 
eigenstates of $\hat M_h$ in the semiclassical limit $h\to 0$. 
To any sequence of eigenstates $\{|\psi_h\rangle\}_{h\to 0}$ corresponds a sequence of 
probability measures on the torus $\{\mu_h\}_{h\to 0}$. 
In the weak-$*$ topology, the set of Borel probability measures
on the torus is compact so the sequence $\{\mu_h\}_{h\to 0}$ admits at least one accumulation
point $\mu$; such a $\mu$
is called a ``semiclassical measure'' (also a ``quantum limit'') of the map $M$, related with
the sequence $\{|\psi_h\rangle\}_{h\to 0}$. 
From Egorov's theorem, the measure
$\mu$ is invariant through the classical map $M$.
A natural question is the following: 

\begin{quote}<<For any $M$-invariant probability measure $\nu$, does there exist a 
sequence of eigenstates of the quantized map $\hat M_h$ admitting $\nu$ as semiclassical measure~?>>
\end{quote}
If the answer is negative, then one wants
to determine the set of semiclassical measures generated by all possible sequences of eigenstates.

If the map $M$ is ergodic (w.r.to the Lebesgue measure on $\t2$), 
it admits a unique absolutely continuous
invariant measure, namely the Lebesgue measure itself 
(which is also the Liouville measure on the symplectic manifold $\t2$). 
On the other hand, each periodic orbit of $M$ supports an invariant Dirac (atomic, pure point) measure. 
As a result, 
if $M$ is an Anosov map (i.e. uniformly hyperbolic on $\t2$),
the space of invariant pure point measures is infinite-dimensional (its closure yields the 
full set of invariant probability measures $\mathfrak{M}_M$ \cite{sigmund}).

We now review some results obtained so far on this issue.
Schnirelman's theorem provides a partial answer to the above question, in the case of an ergodic map: 
``almost all sequences'' of eigenstates admit for
semiclassical measure the Lebesgue measure \cite{Sc,CdV,Z1,BDB}. 
This phenomenon is called ``quantum ergodicity'' in the
mathematics literature. Still, this theorem
does not exclude ``exceptional sequences'' of eigenstates converging to a different semiclassical
measure. 

Extensive numerical studies have shown that many eigenstates of quantum hyperbolic 
systems show an enhanced concentration on one or several (unstable) periodic orbits \cite{heller}. 
Still, it is commonly believed that this
enhancement (called ``scarring'') is weak enough to allow the concerned eigenstates to 
converge (in the weak-$*$ sense) to the Liouville measure. Thus, ``scarring'' must not be mistaken with 
``strong scarring'', that is the existence of a sequence of eigenstates, the
semiclassical measure of which contains a pure point component on some periodic orbit.

``Quantum unique ergodicity'', that is, the absence of any exceptional sequence of eigenstates,
was proven for some families of ergodic linear parabolic maps on $\t2$ \cite{BDB,MR}, using
the fact that for these maps the Lebesgue measure is the unique invariant measure.
On the opposite, a special class of ergodic piecewise affine transformations on $\t2$ have been 
studied and quantized in \cite{schubert}, for which every classical invariant measure is a 
semiclassical measure. In both these cases, the maps are only ``weakly chaotic'', 
in particular they are not mixing and have no periodic point.

In the continuous-time framework, precise
results have been obtained for the eigenstates of the Laplace-Beltrami operator 
on arithmetic manifolds of constant negative curvature in dimension 2 or 3 \cite{RS,LS,PS,K}; the 
corresponding classical dynamics, namely the geodesic flows on the manifolds, are known to be of
Anosov type.
E.~Lindenstrauss \cite{linden}
recently proved quantum unique ergodicity for sequences of joint eigenstates of
the Laplacian and the Hecke operators on arithmetic surfaces
(all eigenstates of the Laplacian are conjectured to be of this type).

In this paper, we restrict ourselves to a very special family of Anosov maps on the
torus, namely the 
linear hyperbolic automorphisms of the 2-torus, also called generalized Arnold's cat maps. 
For any automorphism $A$ of the form $A\equiv {\rm Id}\bmod{4}$ and any value of $h$,  Rudnick and 
Kurlberg have defined a family of ``Hecke operators'' commuting with the quantum map $\hat A_h$, 
and proven unique quantum ergodicity
for the sequences of joint eigenstates \cite{kuru1}. However, as opposed to the case of 
arithmetic surfaces, many eigenstates of $\hat A_h$ are not Hecke eigenstates, leaving open the
possibility of exceptional sequences.
In \cite{debrievre02}, Bonechi and De Bièvre have
shown that for any automorphism $A$, a semiclassical measure of $A$ cannot 
be completely localized (or completely ``scarred''), in that
the weight of its pure point
component is bounded above by $\left(\sqrt{5}-1\right)/2\simeq 0.62$
(their proof, which assumes this component to be supported on \emph{finitely many} periodic 
points, also applies to ergodic automorphisms on higher-dimensional symplectic tori). 
In \cite{fred-steph02}, sequences of eigenstates of $\hat A_h$ were explicitly constructed, for which
the semiclassical measure has a pure point component of weight $1/2$ localized on a finite
set of periodic orbits, the remaining part of the measure being Lebesgue.
In this paper we improve the results of \cite{debrievre02} as follows:
\pagebreak
\begin{thm}\label{thm:restriction}
Let $A$ be a hyperbolic automorphism of $\t2$, and $\mu$ be a normalized semiclassical measure of 
$A$. Splitting
$\mu$ into its pure point, Lebesgue and singular continuous components, 
$\mu=\mu_{pp}+\mu_{Leb}+\mu_{sc}$, the following inequalities hold between the weigths of these
components: $\mu_{Leb}(\t2)\geq \mu_{pp}(\t2)$, which implies  
$\mu_{pp}(\t2)\leq 1/2$.
\end{thm}
The states constructed in \cite{fred-steph02} saturate this upper bound: 
they are ``maximally scarred''.

After recalling the definition of the quantized automorphisms (Section \ref{s:QHA}), we
prove in Section~\ref{s:evolution} two ``dynamical'' propositions 
(the first one was proven in \cite{debrievre02} in a more general context). 
They both deal with the 
evolution through $\hat A_h$ up to the ``Ehrenfest time''
$T\sim |\log h|/\lambda$, of quantum states with prescribed initial localization properties. 
Using these propositions, we then show in 
Section~\ref{s:maximal_loc} that for any finite union $\S$ of periodic orbits, any
semiclassical measure $\mu$ of $A$ satisfies $\mu(\S)\leq \mu_{Leb}(\t2)$ 
(Theorem~\ref{thm:maximal_measure}), 
from where the above theorem is a straightforward corollary. 
In final remarks, we draw consequences of the above theorem, concerning the
determination of the set $\mathfrak{M}_{A,{\rm SC}}$ of semiclassical measures for 
the automorphism $A$. We also discuss
possible extensions of these results to a broader class of 
Anosov systems.

%%%%%%%%%%%%%%%%%%%%%%%%%%%%%%%%%%%%%%%%%%%%%%%%%%%%%%%%%%%%%%%%%%%%%%%%%%%%%%%
%%%%%%%%%%%%%%%%%%%%%%%%%%%%%%%%%%%%%%%%%%%%%%%%%%%%%%%%%%%%%%%%%%%%%%%%%%%%%%%

\section{Quantum hyperbolic automorphisms on $\t2$\label{s:QHA}}

\subsection{Quantum mechanics on $\t2$}
We briefly describe the quantum mechanics on the 2-torus phase space as defined in \cite{hannay,DE}. 
The Hilbert space of quantum 
states corresponding to Planck's constant $h$ will be called $\hh$: quantum states can be
identified with distributions
$\psi(q)\in\S'(\R)$ such that $\psi(q+1)=\psi(q)$ and $(\mathcal{F}\psi)(k+h^{-1})=
(\mathcal{F}\psi)(k)$,
where $\mathcal{F}$ is the Fourier transform.  $\hh$ is a nontrivial vector space
iff $h^{-1}\in\N^*$, and then $\hh\simeq \C^{h^{-1}}$. 
In what follows, $h$ will always be taken of that form; 
the semiclassical limit is therefore 
defined as the limit $h\to 0$, $h^{-1}\in\N$.

For any classical observable $f\in C^\infty(\t2)$, we note respectively 
$\hat f=Op_h(f)$ its Weyl quantization and $\hat f^{\AW}=Op_h^{\AW}(f)$ 
its anti-Wick quantization on $\hh$ \cite{BDB}. The anti-Wick quantized operator satisfies the
property $\Vert \hat f^{\AW}\Vert\leq \Vert f\Vert_\infty=\sup_{\t2}(|f|)$.
To any state $|\psi_h\rangle \in \hh$ correspond the Wigner and Husimi measures $\tilde\mu_h$,
$\mu_h$ defined as \cite{BDB}
\begin{align}
\tilde\mu_h(f)&=\int_{\t2}dx\,W_{\psi_h}(x)\,f(x)=\langle \psi_h|\hat{f}|\psi_h\rangle,\\
\mu_h(f)&=\int_{\t2}dx\,H_{\psi_h}(x)\,f(x)=\int_{\t2}\frac{dx}{h}\abs{\langle\psi_h|x\rangle}^2\,f(x)
=\langle\psi_h|\hat{f}^{\AW}|\psi_h\rangle.
\end{align}
The ket $|x\rangle$ denotes the (asymptotically
normalized) coherent state in $\hh$ localized at the point $x=(q,p)$, with widths
$\Delta q\sim\Delta p\sim \sqrt{\hbar/2}$.
While the Husimi density $H_{\psi_h}(x)$ is a non-negative smooth function on $\t2$, 
the ``Wigner function'' $W_{\psi_h}(x)$ is a finite
combination of delta peaks with real (possibly negative) coefficients.

We now consider an infinite sequence of states $\{|\psi_h\rangle\in\hh\}_{h\to 0}$. By definition, 
the corresponding sequence
of Husimi measures $\{\mu_h\}$ weak-$*$ converges to 
$\mu$ iff for any smooth observable $f$, one has
$\mu_h(f)\hto0\mu(f)$.
The sequence of signed measures $\{\tilde\mu_h\}$ then admits the same weak-$*$
limit ($\mu_h$ is the convolution of $\tilde\mu_h$ by a Gaussian kernel of width $\sim\sqrt{\hbar}$).
The limit (Borel) measure $\mu$ is then called the 
\textbf{semiclassical measure} of the sequence $\{|\psi_h\rangle\}_{h\to 0}$ (by a slight abuse of 
language, we also say that the sequence of states $\{|\psi_h\rangle\}_{h\to 0}$ converges to $\mu$). 

If the states $|\psi_h\rangle$ are (normalized) eigenstates of a quantized map $\hat M_h$, then $\mu$ is 
called a \textbf{semiclassical measure for $M$}.
In that case, Egorov's property, that is the semiclassical commutation of time evolution
and quantization:
\begin{equation}\label{egorov}
\forall t\in\Z,\quad \norm{\hat M_h^t\, Op_h(f)\hat M^{-t}_h-Op_h(f\circ M^t)}\hto0 0,
\end{equation}
implies that $\mu$ is an invariant measure for the classical map $M$.

A sequence of states $\{|\psi_h\rangle\}_{h\to 0}$ is said to be \textbf{equidistributed}
if it converges semiclassically to (a multiple of) the Lebesgue measure
on $\t2$ (noted $dx$), i.e. iff for a certain $C>0$ and for any
observable $f$, $\mu_h(f)$ (equivalently $\tilde\mu_h(f)$)
converges to $C\int_{\t2} dx\,f(x)$.
On the opposite, 
a sequence $\{|\psi_h\rangle\}_{h\to 0}$ is called \textbf{localized}
iff it converges to a pure point measure on $\t2$, that is if
there exists a countable set of points $\{x_i\}$ and weights 
$\alpha_i>0$, $\sum_i\alpha_i<\infty$
such that for any observable $f$,
$\mu_h(f)\hto0 \sum_i \alpha_i f(x_i)$. 

%%%%%%%%%%%%%%%%
\subsection{Quantum hyperbolic automorphisms}

An automorphism of $\t2$, or generalized cat map,
is the diffeomorphism on $\t2$ defined by the action
of a matrix $A\in {\rm SL}(2,\Z)$ on the point
$x=\binom{q}{p}\in \t2$. The map itself will also be denoted by $A$. It is uniformly hyperbolic
on $\t2$ (therefore of Anosov type) iff $|\tr(A)|>2$.
Depending on the sign of the trace, the
eigenvalues of $A$ are of the form $\{\pm \e^\lambda,\,\pm \e^{-\lambda}\}$, where $\lambda>0$
is the Lyapounov exponent. The corresponding eigenaxes define the unstable/stable directions at
any point $x\in\t2$, and their
projections on
the torus make up the unstable and stable manifolds
of the origin (which is an obvious fixed point).
We will use the property that the slopes of these axes
are irrational, so that both manifolds are dense on $\t2$.

Under the condition 
$A\equiv \begin{pmatrix}1&0\\0&1\end{pmatrix}$ or $\begin{pmatrix}0&1\\1&0\end{pmatrix}\bmod 2$, 
the linear automorphism $A$ can be quantized on any $\hh$, yielding a unitary matrix 
$\hat A_h$ of dimension $h^{-1}$ \cite{hannay,DE}.
For simplicity, we will assume in the following sections that $A$ is of that form. Yet, 
this restriction can easily be lifted:
any matrix $A\in{\rm SL}(2,\Z)$ can be quantized on $\hh$ if we restrict 
$h^{-1}$ to take $even$ values, or extend
the definition of $\hh$ to allow nontrivial ``Bloch angles'' \cite{DE,BDB}. All our results can be
straightforwardly generalized to those cases.

Let us call $T_{\mathbf{v}}$ the classical translation
by the vector $\mathbf{v}\in \R^2$. It can be naturally quantized as a unitary matrix
$\hat{T}_{\mathbf{v}}$ on $\hh$ iff $\mathbf{v}$ belongs to the square lattice $(h\Z)^2$, that is  
iff $\mathbf{v}=h\k $ for a certain $\k \in\Z^2$. The operator
$\hat{T}_{h\k }$ 
can also be interpreted as the Weyl quantization $Op_h(F_\k)$ of the complex-valued observable
$F_{\k }(q,p)=\exp \big(-2\pi \i(k_1 p-k_2 q)\big)$
on $\t2$. 

For any $h$, the following intertwining relation holds between the quantum automorphism 
$\hat A_h$ and the quantized translations:
\begin{equation}\label{eq:linearity}
\forall\k \in\Z^2,\qquad 
\hat{A}_h\hat{T}_{h\k }\hat{A}^{-1}_h=\hat{T}_{hA\k }\Longleftrightarrow
\hat{A}_h\,Op_h(F_\k )\,\hat{A}^{-1}_h=Op_h(F_\k \circ A^{-1}).
\end{equation}
Comparing with Eq.~\eqref{egorov}, we see that the above identity realizes an {\it exact} 
Egorov property: 
quantization exactly commutes with time evolution, for arbitrary times $t$ and arbitrary $h$ \cite{hannay}. 
This exact equality is \emph{characteristic of linear maps}, and will be crucial
in the next sections.

%%%%%%%%%%%%%%%%%%%%%%%%%%%%%%%%%%%%%%%%%%%%%%%%%%%%%%%%%%%%%%%%%%%%%%%%%%%%%%%
%%%%%%%%%%%%%%%%%%%%%%%%%%%%%%%%%%%%%%%%%%%%%%%%%%%%%%%%%%%%%%%%%%%%%%%%%%%%%%%

\section{Localized states evolve into equidistributed states\label{s:evolution}}
We define the Ehrenfest time or log-time corresponding to the quantum map $\hat A_h$ by 
$$
T=\left[\frac{\left|\log h\right|}{\lambda }\right],
$$
where $[.]$ denotes the integer part (we will always omit to indicate the $h$-dependence of $T$).

Consider a set of $n$ points $\S=\left\{x_i\right\}_{i=1\rightarrow n}$ on $\t2$,
such that for a certain integer $d>0$, all
vectors $(x_i-x_j)$ belong to the lattice $\frac{1}{d}\mathbb{Z}^{2}$.
Assign to each point $x_i$ a weight $\alpha_i>0$, with $\sum_i\alpha_i=1$, 
and define the Dirac probability measure 
$\delta_{\S,\alpha}=\sum_{i=1}^n\alpha_i\,\delta_{x_i}$ on the torus. 

\begin{pro}
\label{prop:evolution}Suppose that a sequence of states 
$\{|\varphi_{h}\rangle \in \mathcal{H}_{h}\}_{h\to 0}$
converges to the measure $\delta_{\S,\alpha}$.
Then the sequence of states $\{|\varphi'_h \rangle =\hat{A}_h^{T}|\varphi_h\rangle\}$
is semiclassically equidistributed.
\end{pro}
This property has been proven in a more general setting (higher
dimensional automorphisms, time $T$ replaced by a time interval) in \cite[Theorem 5.1]{debrievre02}.
To be self-contained, we give below a ``fast'' proof of this proposition for our case.
\begin{proof}
As a first step, we draw consequences from the assumption 
$\mu_h\htoo \delta_{\S,\alpha}$, where $\mu_h$ is the Husimi measure of $|\varphi_h\rangle$. 
Denoting by $D(x_i,a)$ the disk of radius $a>0$ centered at $x_i$, and by 
$D(\S,a)\defi\cup_{i=1}^n D(x_i,a)$ the corresponding neighbourhood of $\S$,
the weak-$*$ convergence of $\mu_h$ implies that $\mu_h(\t2)\htoo 1$ and that for any $a>0$, 
$\mu_h\big(\t2\!\setminus\! D(\S,a)\big)\htoo 0$. From a standard diagonal argument, one can 
then construct a decreasing sequence $a_h>0$, $a_h\htoo 0$ such that
\begin{equation}\label{e:local}
\lvert\mu_h\big(\t2\big)-1\rvert\leq a_h\quad\mbox{and}
\quad\mu_h\big(\t2\setminus D(\S,a_h)\big)\leq a_h.
\end{equation}

In a second step, we remark that proving the equidistribution of the sequence 
$\left\{|\varphi'_{h}\rangle\right\}$ amounts to prove that 
for any fixed $\k \in \Z^2$, 
\begin{equation}\label{e:equidistribution}
\tilde\mu'_{h}(F_{\k})=\langle\varphi'_h|\hat{T}_{h\k }|\varphi'_h\rangle 
\hto0\int_{\t2}F_{\k}(x)\;dx=\delta_{\k ,\mathbf{0}}.
\end{equation}
The case $\k=\mathbf{0}$ is obvious since $\tilde\mu'_h(\t2)=\tilde\mu_h(\t2)=\mu_h(\t2)\htoo 1$ 
by assumption. 
We now select a fixed wave vector $\mathbf{0}\neq\k \in\Z^2$, and 
study the above LHS. The intertwining relation 
(\ref{eq:linearity}) allows us to rewrite it as 
\begin{equation}\label{eq:trick}
\langle \varphi'_h|\hat{T}_{h\k }|\varphi'_h\rangle 
=\langle \varphi_h|\hat A_h^{-T}\hat{T}_{h\k }\hat A_h^T|\varphi_h\rangle 
=\langle \varphi_h|\hat{T}_{h\k '_h}|\varphi_h\rangle,
\end{equation}
with the vector $\k '_h=A^{-T}\k $.
From the definition of the Ehrenfest time, the `large' eigenvalue of $A^{-T}$ is 
$(\pm e^{\lambda})^T=\pm C(h)\frac{1}{h}$, where
the prefactor satisfies $e^{-\lambda }\leq C(h)\leq 1$. The decomposition of that vector
in the eigenbasis of $A$ then reads:
\begin{equation}\label{e:hk'_decompo}
h\k'_h=\pm e^{\lambda T}h\k ^{stable}\pm e^{-\lambda T}h\k ^{unstable}
=\pm C(h)\k ^{stable}+\O(h^2).
\end{equation}
The vector $h\k '_h$ thus has a finite length (bounded from above and from below uniformly in $h$), 
and is asymptotically parallel to the stable axis.
Because the slope of that axis is irrational, the {\it distance}
between $h\k '_h$ and the lattice $\frac{1}{d}\Z^2$
is bounded from below by a constant $c(\k)>0$ uniformly w.r.to $h$. 

To finish the proof, we write the above overlap as
a coherent state integral:
\begin{equation}\label{e:coh_st}
\abs{\langle \varphi_h|\hat{T}_{h\k _h'}|\varphi_h\rangle}
=\left\vert\int_{\t2} \frac{dx}{h}\;\langle\varphi_h|x\rangle
\langle x|\hat{T}_{h\k _h'}|\varphi_h\rangle\right\vert
\leq\int_{\t2}\frac{dx}{h}\;\abs{\langle x|\varphi_h\rangle}\:\abs{\langle x-h\k_h'|\varphi_h\rangle},
\end{equation}
We then split the integral on the RHS between $D(\S,a_h)$
and its complement. The integral on $D(\S,a_h)$
is estimated through a Cauchy-Schwarz inequality:
$$
\int_{D(\S,a_h)}\frac{dx}{h}\;\abs{\langle x|\varphi_h\rangle}\:\abs{\langle x-h\k_h'|\varphi_h\rangle}
\leq \sqrt{\mu_h\big(D(\S,a_h)\big)\;\mu_h\big(D(\S,a_h)-h\k '_h\big)}.
$$
Due to above-mentioned property of $h\k '_h$ and the fact that $x_i-x_j\in\frac{1}{d}\Z^2$, 
for small enough $a_h$ the set 
$D(\S,a_h)-h\k '_h$ has no intersection with $D(\S,a_h)$. As a consequence, using 
\eqref{e:local}, the second factor on the RHS
is bounded above by $\sqrt{a_h}$, and the full RHS by $\sqrt{a_h(1+a_h)}$.
The remaining integral over $\t2\!\setminus\! D(\S,a_h)$ admits the same upper bound for similar
reasons.
\end{proof}

\begin{pro}\label{pro:crossed}
Let $(\S,\alpha)$ be a finite weighted set of $A$-periodic points, and $\nu$ an 
$A$-invariant probability measure satisfying $\nu(\S)=0$.
Suppose that the sequence 
$\{|\varphi_{\S,h}\rangle\}_{h\to 0}$
converges semiclassically to the measure $\delta_{\S,\alpha}$, and that a second sequence 
$\{|\varphi_{\nu,h}\rangle\}_{h\to 0}$ converges to $\nu$. 
Then the states $|\varphi'_{\S,h} \rangle =\hat{A}_h^{T}|\varphi_{\S,h}\rangle$, 
$|\varphi'_{\nu,h} \rangle =\hat{A}_h^{T}|\varphi_{\nu,h}\rangle$ satisfy:
$$
\forall \k\in\Z^2,\quad \langle\varphi'_{\S,h}|\hat T_{h\k}|\varphi'_{\nu,h}\rangle\hto0 0. 
$$
\end{pro}
\begin{proof}
We use similar methods as for the previous proposition. Namely, as in Eq.~\eqref{eq:trick} 
we rewrite the overlap as
\begin{equation}\label{trick2}
\langle\varphi'_{\S,h}|\hat T_{h\k}|\varphi'_{\nu,h}\rangle
=\langle\varphi_{\S,h}|\hat T_{h\k '_h}|\varphi_{\nu,h}\rangle
=\int_{\t2}\frac{dx}{h}\;\langle\varphi_{\S,h}|x\rangle\langle x
|\hat{T}_{h\k _h'}|\varphi_{\nu,h}\rangle
\end{equation}
with $h\k '_h$ given by Eq.~\eqref{e:hk'_decompo}. 
We want to cut this integral into two parts. Using the same notations as above, 
the assumptions of the proposition imply the existence of 
a decreasing sequence $a_h\hto0 0$ such that the Husimi mesures of $\varphi_{\S,h}$ and 
$\varphi_{\nu,h}$ satisfy 
$$
\mu_{\S,h}\big(\t2\setminus D(\S,a_h)\big)\leq a_h,\quad \mu_{\S,h}\big(\t2\big)\leq 1+a_h
\quad\mbox{and}\quad\mu_{\nu,h}\big(\t2\big)\leq 1+a_h.
$$
These inequalities also hold if we replace $a_h$ by any decreasing sequence $b_h\geq a_h$, $b_h\to 0$.
We can now bound the part of the integral \eqref{trick2} over the set $\t2\!\setminus\! D(\S,b_h)$:
\begin{equation*}
\begin{split}
\left\vert\int_{\t2\setminus D(\S,b_h)}
\frac{dx}{h}\;\langle\varphi_{\S,h}|x\rangle\langle x|
\hat{T}_{h\k _h'}|\varphi_{\nu,h}\rangle\right\vert
&\leq \sqrt{\mu_{\S,h}\big(\t2\!\setminus\! D(\S,b_h)\big)\;\mu_{\nu,h}\big(\t2\big)}\\
&\leq \sqrt{b_h(1+b_h)}\hto0 0.
\end{split}
\end{equation*}
The second part of the integral is also estimated by Cauchy-Schwarz:
$$
\left\vert\int_{D(\S,b_h)}
\frac{dx}{h}\;\langle\varphi_{\S,h}|x\rangle\langle x|\hat{T}_{h\k _h'}
|\varphi_{\nu,h}\rangle\right\vert
\leq\sqrt{\mu_{\S,h}\big(\t2\big)\;\mu_{\nu,h}\big(D(\S,b_h)-h\k'_h\big)}.
$$
We now show that the sequence $b_h$ can be chosen such that the second factor under the
square root vanishes in the semiclassical limit, so that the full RHS vanishes as well.
Indeed, using
Eq.~\eqref{e:hk'_decompo} and the bound on $C(h)$, one realizes that
for $h$ small enough, the set $D(\S,b_h)-h\k'_h$ is 
contained in the thin ``tube'' 
$$
T_{b_h}\defi D(\S,2b_h)+\big([-1,-\e^{-\lambda}]\cup[e^{-\lambda},1]\big)\k ^{stable}.
$$ 
As $b_h\to 0$, this tube
decreases to 
$T_0=\S+\big([-1,-\e^{-\lambda}]\cup[e^{-\lambda},1]\big)\k ^{stable}$, that is a union of segments
of the stable manifold. 
The following simple lemma is proven in Appendix~\ref{a:stable}.
\begin{lem}\label{lem:nu(stable)}
For any invariant probability measure $\nu$, any finite set of periodic points $\S$ and
any $0<a<b<\infty$, one has
$\nu(\S+[a,b]\mathbf{e}^{stable})=0$, where $\mathbf{e}^{stable}$ is a unit vector in
the stable direction.
\end{lem} 
\noindent This lemma implies that $\nu(T_0)=0$.
Since $b_h\searrow 0$, the regularity of the Borel measure $\nu$ entails:
$$
\lim_{b_h\to 0}\nu(T_{b_h})=\nu(T_0)=0.
$$
Since by assumption the Husimi measures $\mu_{\nu,h}$ converge
towards $\nu$, one can by a diagonal argument choose the sequence $b_h\searrow 0$ 
(making sure that $b_h\geq a_h$) such that:
$$
\mu_{\nu,h}\big(T_{b_h}\big)\hto0 0.
$$ 
This finally implies that $\mu_{\nu,h}\big(D(\S,b_h)-h\k'_h\big)\hto0 0$.
\end{proof}

%%%%%%%%%%%%%%%%%%%%%%%%%%%%%%%%%%%%%%%%%%%%%%%%%%%%%%%%%%%%%%%%%%%%%%%%%%%%%%%
%%%%%%%%%%%%%%%%%%%%%%%%%%%%%%%%%%%%%%%%%%%%%%%%%%%%%%%%%%%%%%%%%%%%%%%%%%%%%%%

\section{Maximal localization of the semiclassical measures\label{s:maximal_loc}}

We call
$\tau$ a periodic orbit of $A$ of period $T_\tau$,
and denote its associated measure by
$$
\delta_\tau=\frac{1}{T_\tau}\sum_{x\in\tau}\delta_{x}.
$$
We consider a {\it finite} set $\mathcal{F}$ of periodic
orbits, associate a weight $w_\tau>0$ to each orbit
($\sum_{\tau\in\mathcal{F}}w_\tau=1$), and construct 
the pure point invariant measure 
$$
\delta_{\mathcal{F},w}=\sum_{\tau\in\mathcal{F}} w_\tau\,\delta_{\tau}.
$$
All periodic points of $A$ have rational coordinates, so this measure is a particular instance
of the measure $\delta_{\S,\alpha}$ considered in Proposition~\ref{prop:evolution}.
Indeed, grouping the periodic orbits of $\mathcal{F}$ together yields the set of rational points
$\S=\S_{\mathcal{F}}=\cup_{\tau\in\mathcal{F}}\tau=\{x_1,\ldots,x_n\}$, and the
weight allocated to each point $x_i$ reads
$\alpha_i=\frac{w_\tau}{T_\tau}$. From now on, the two notations
$\delta_{\mathcal{F},w}$ and $\delta_{\S,\alpha}$ will refer to the same invariant measure.
We now state the central result of this article.
\begin{thm}
\label{thm:maximal_measure}Let $\mu$ be a normalized $A$-invariant Borel measure of $A$, and
$(\mathcal{F},w)$ a finite weighted set of periodic orbits. $\mu$ may be 
decomposed into $\mu =\beta\delta_{\mathcal{F},w}+(1-\beta)\nu$, where
$\nu$ is a normalized invariant measure satisfying  $\nu(\S_{\mathcal{F}})=0$.
If $\mu$ is a semiclassical measure of $A$, then its Lebesgue component
has a weight $\geq\beta$, which in turn implies $\beta\leq 1/2$.
\end{thm}
Any invariant Borel measure $\mu\in\mathfrak{M}_A$ 
can obviously be decomposed in the above way, with $0\leq\beta\leq 1$.
We do not assume the measure $\nu$ to
be continuous, but allow it to contain a pure point component localized on a (possibly countable)
set of periodic orbits disjoint with $\mathcal{F}$.
The statement of the theorem is of course stronger if we include in $(\mathcal{F},w)$ as many 
orbits as possible. By a simple limit argument, one may eventually take for $\nu$ the
continuous component
of $\mu$, allowing $(\mathcal{F},w)$ to be a {\it countable} weighted set of orbits
(still taking $\sum_{\tau\in\mathcal{F}}w_\tau=1$): one then obtains
Theorem~\ref{thm:restriction} as a simple corollary of the one above.

\medskip

\begin{proof}[Proof of Theorem \ref{thm:maximal_measure}:]
There are two steps in the proof. Let $\{|\psi_h\rangle\}_{h\to 0}$ be a sequence 
of eigenstates of $\hat A_h$ admitting $\mu$ as semiclassical measure.
Our first objective is to decompose
the state $|\psi_h\rangle$ into $|\psi_{\S,h}\rangle +|\psi_{\nu,h}\rangle$,
such that the sequence $\{|\psi_{\S,h}\rangle\}$ (resp. $\{|\psi_{\nu ,h}\rangle\}$)
converges to the measure $\beta\delta_{\S,\alpha}$ (resp. the measure $(1-\beta)\nu$).
This decomposition will be obtained by ``projecting''
$|\psi_h\rangle $ on appropriate ($h$-dependent) small neighboorhoods of $\S$.

In the second part we will be guided by the following simple idea.
Because $|\psi_h\rangle $ is an eigenstate of $\hat A_h$, 
$|\psi_h\rangle \propto \hat{A}_h^t|\psi_h\rangle =\hat{A}_h^t|\psi_{\S,h}\rangle 
+\hat{A}_h^t|\psi_{\nu,h}\rangle $ for any $t\in\Z$, in particular for the Eherenfest time $t=T$.
From Proposition \ref{prop:evolution}, the sequence of states 
$\{\hat{A}_h^T|\psi_{\S,h}\rangle\}$
is equidistributed; together with Proposition \ref{pro:crossed}, that implies
that the semiclassical measure of $\hat{A}_h^T|\psi_h\rangle$ (that is, $\mu$) contains a Lebesgue
part of weight $\geq\beta$. 

%%%%%%%%%%%%%%%%%%%%%%%%%%%%%%%%%%%%%%%%%%%%%%%%%%%%%%%%%%%%%%%%%%%%%

\subsection*{First part: splitting the eigenstates}
To ``project'' the eigenstates in a small neighbourhood of $\S$, we will use 
a function $\vartheta\in C^\infty(\R^2)$ satisfying $0\leq\vartheta(x)\leq 1$ on $\R^2$,
$\vartheta(x)=0$ outside the disk
$D(0,2)$ and $\vartheta(x)=1$ inside the disk $D(0,1)$.  

For any (small) $r>0$, and any point $x_i\in\S$, let 
$\vartheta_{i,r}(x)=\sum_{\mathbf{n}\in\Z^2}\vartheta\big(\frac{x-x_i-n}{r}\big)$, which can
be seen as a smooth function on $\t2$, localized around $x_i$. We will 
also need the function $\vartheta_r=\sum_{i=1}^n \vartheta_{i,r}$ to take all points of $\S$
into account: this function
is supported in the neighbourhood $D(\S,2r)$ of $\S$. There is an $r_0>0$ such that
the disks $D(x_i,2r_0)$ do not overlap each other, which implies $\vartheta_{i,r}\vartheta_{j,r}\equiv 0$ 
if $i\neq j$.

Fixing $0<r\leq r_0$, we consider the anti-Wick quantization of these functions, 
$\hat\vartheta_{i,r}^{\AW}$ and use them to decompose $|\psi_h\rangle$.
We first observe that for any $x_i\in\S$,
$$
\langle\psi_h|\hat\vartheta_{i,r}^{\AW}|\psi_h\rangle \hto0
\beta \delta_{\S,\alpha}(\vartheta_{i,r})+(1-\beta)\nu(\vartheta_{i,r})
=\beta\alpha_i +(1-\beta)\nu(\vartheta_{i,r}).
$$
Furthermore, one has for any pair $x_i,\,x_j\in\S$,
$$
\langle\psi_h|\hat\vartheta_{i,r}^{\AW}\hat\vartheta_{j,r}^{\AW}|\psi_h\rangle \hto0
\beta \delta_{\S,\alpha}(\vartheta_{i,r}\vartheta_{j,r}) +
(1-\beta)\nu(\vartheta_{i,r}\vartheta_{j,r})
=\delta_{ij}\left(\beta\alpha_i +(1-\beta)\nu(\vartheta_{i,r}^2)\right).
$$
On the other hand, the regularity of the (Borel) measure $\nu$ entails:
$$
\forall x_i\in\S,\qquad\nu(\vartheta_{i,r})\rto0 \nu(\S)=0\qquad\mbox{and similarly}
\qquad \nu(\vartheta_{i,r}^2)\rto0 0.
$$
From the above limits, one can
construct by a diagonal argument a decreasing sequence of radii $r(h)\htoo 0$ such that
\begin{equation}\label{microloc}
\forall x_i,\;x_j\in\S,
\quad\langle\psi_{h}|\hat\vartheta_{i,r(h)}^{\AW}|\psi_h\rangle \hto0\beta\alpha_i \quad
\mbox{and}
\quad
\langle\psi_h|\hat\vartheta_{i,r(h)}^{\AW}\hat\vartheta_{j,r(h)}^{\AW}|\psi_h\rangle\hto0 
\delta_{ij}\beta\alpha_i.
\end{equation} 
We now show that the two sequences of vectors
\begin{equation}
|\psi_{\S,h}\rangle\defi\hat\vartheta_{r(h)}^{\AW}|\psi_h\rangle ,\qquad 
|\psi_{\nu,h}\rangle\defi\big({\rm Id}-\hat\vartheta_{r(h)}^{\AW}\big)|\psi_h\rangle
\end{equation}
provide the desired decomposition of $|\psi_h\rangle$, that is the corresponding
sequences respectively 
converge to the measures $\beta\delta_{\S,\alpha}$ and $(1-\beta)\nu$.
The first statement $\mu_{\S,h}\htoo \beta\delta_{\S,\alpha}$ seems quite natural: 
the operator $\hat\vartheta_{r(h)}^{\AW}$ acts as a ``microlocal  projector''
onto the set $\S$. This property is precisely expressed in the following lemma, which we
prove in Appendix~\ref{a:obs-micro}.
\begin{lem}\label{lem:observable-microloc}
For any
$f\in C^\infty(\t2)$ and for any $x_i\in\S$, one has
$$
\norm{\hat f^{\AW}\,\hat\vartheta_{i,r(h)}^{\AW}- f(x_i)\,\hat\vartheta_{i,r(h)}^{\AW}}
\hto0 0,
$$
where $\norm{.}$ is the operator norm on $\hh$.
\end{lem}
\noindent This lemma together with the properties \eqref{microloc} immediately yield the
required limits:
\begin{align}
\forall f\in C^\infty(\t2),\qquad \langle\psi_{\S,h}|\hat f^{\AW}|\psi_{\S,h}\rangle
&\hto0\sum_{i}\beta\alpha_i f(x_i)=\beta\delta_{\S,\alpha}(f),\label{e:psi_S}\\
\langle \psi_{\nu,h}|\hat f^{\AW}|\psi_{\nu,h}\rangle&\hto0 \mu(f)-\beta\delta_{\S,\alpha}(f)
=(1-\beta)\nu(f).\label{e:psi_nu}
\end{align}

%%%%%%%%%%%%%%%%%%%%%%%%%%%%%%%%%%%%%%%%%%%%%%%%%%%%%%%%%%%%%%%%%%%%%%%%%%%%%%%
\medskip

\subsection*{Second part: playing with time evolution}
We will follow a strategy 
close to the one used to prove Proposition~\ref{prop:evolution}. We consider a
fixed $\k\in\Z^2$, and focus on
the overlap $\langle\psi_h|\hat T_{h\k}|\psi_h\rangle$. 
From the semiclassical assumption on 
$\{|\psi_h\rangle\}$, one has
\begin{equation}\label{assumption}
\langle\psi_h|\hat T_{h\k}|\psi_h\rangle\hto0 \mu(F_{\k})=\beta\delta_{\S,\alpha}(F_{\k})+
(1-\beta)\nu(F_{\k}).
\end{equation}
On the other hand, since $|\psi_h\rangle=|\psi_{\S,h}\rangle+|\psi_{\nu,h}\rangle$ 
is an eigenstate of $\hat A_h$, one may rewrite
\begin{align}
\langle\psi_h|\hat T_{h\k}|\psi_h\rangle
&=\langle\psi_h|\hat A_h^{-T}\,\hat T_{h\k}\,\hat A_h^T|\psi_h\rangle\\
&=\langle\psi'_{\S,h}|\hat T_{h\k}|\psi'_{\S,h}\rangle+
2\Re\left(\langle\psi'_{\nu,h}|\hat T_{h\k}|\psi'_{\S,h}\rangle\right)
+\langle\psi'_{\nu,h}|\hat T_{h\k}|\psi'_{\nu,h}\rangle,
\label{e:3terms}
\end{align}
with the notation $|\psi'_{\S,h}\rangle=\hat A_h^T|\psi_{\S,h}\rangle$, 
$|\psi'_{\nu,h}\rangle=\hat A_h^T|\psi_{\nu,h}\rangle$ and $T$ is the Ehrenfest time. 
We are now in position to collect the dynamical results of Section~\ref{s:evolution}:
\begin{itemize}
\item From the asymptotic localization \eqref{e:psi_S} of $|\psi_{\S,h}\rangle$ and 
Proposition~\ref{prop:evolution}, 
the first term in \eqref{e:3terms} converges to $\beta dx(F_{\k})=\beta\delta_{\k,\mathbf{0}}$ as 
$h\to 0$.
\item From the properties (\ref{e:psi_S}-\ref{e:psi_nu}) and Proposition~\ref{pro:crossed}, 
the cross-terms 
in \eqref{e:3terms} vanish in the semiclassical limit:
$\langle\psi'_{\nu,h}|\hat T_{h\k}|\psi'_{\S,h}\rangle\hto0 0$.
\end{itemize}
Using Eq.~\eqref{assumption}, the last term in \eqref{e:3terms} has therefore the semiclassical behaviour
$$
\langle\psi'_{\nu,h}|Op_h(F_{\k})|\psi'_{\nu,h}\rangle\hto0 \mu(F_{\k})-\beta dx(F_{\k}).
$$
Since this limit holds for any $\k\in\Z^2$, it means that the the sequence 
$\{|\psi'_{\nu,h}\rangle\}_{h\to 0}$ admits the semiclassical measure $\mu-\beta dx$.
Because semiclassical measures are limits of Husimi measures, they cannot contain any negative
part. Therefore, $\mu-\beta dx$ must be a positive measure, which 
implies that the Legesgue component of $\mu$  has a weight $\geq \beta$. This 
component being contained in $(1-\beta)\nu$, one has finally 
$(1-\beta)\geq \beta\Leftrightarrow\beta\leq 1/2$.
\end{proof}

%%%%%%%%%%%%%%%%%%%%%%%%%%%%%%%%%%%%%%%%%%%%%%%%%%%%%%%%%%%%%%%%%%%%%%%%%%%%%
%%%%%%%%%%%%%%%%%%%%%%%%%%%%%%%%%%%%%%%%%%%%%%%%%%%%%%%%%%%%%%%%%%%%%%%%%%%%%

\section{Remarks and comments}

\subsection{On the set of semiclassical measures}
Theorem~\ref{thm:restriction} constrains the set of semiclassical measures $\mathfrak{M}_{A,{\rm SC}}$
to be a \emph{proper subset} of the set $\mathfrak{M}_A$ of invariant Borel measures. 
One can easily show that 
$\mathfrak{M}_{A,{\rm SC}}$ is a \emph{closed set} of $\mathfrak{M}_A$. 
Indeed, if for any $n>0$ the sequence
$S_n=\{|\psi_{h,n}\rangle\}_{h\to 0}$ converges towards a normalized
semiclassical measure $\mu_n$, and that in turn
the measures $\mu_n$ weak-$*$ converge to a measure $\mu$, then one can extract a function 
$n(h)\htoo \infty$ such that
$\{|\psi_{h,n(h)}\rangle\}_{h\to 0}$ converges to $\mu$.

Every open neighbourhood of  $\mathfrak{M}_A$ contains
a pure point measure of type $\delta_\tau$ ($\tau$ a periodic orbit) \cite{sigmund,marcus}, 
therefore Theorem~\ref{thm:restriction} implies that the set $\mathfrak{M}_{A,{\rm SC}}$
is \emph{nowhere dense} in $\mathfrak{M}_A$ ({\it i.e.} its interior is empty).

On the other hand, the 
results of \cite{fred-steph02} show that $\mathfrak{M}_{A,{\rm SC}}$ contains all measures
of the type $\frac{\delta_\tau+dx}{2}$. Since the measures $\{\delta_\tau\}$ are dense in 
$\mathfrak{M}_A$ and the set $\mathfrak{M}_{A,{\rm SC}}$ is closed, this implies
$$
\forall \nu\in\mathfrak{M}_A,\qquad \frac{\nu+dx}{2}\in\mathfrak{M}_{A,{\rm SC}}.
$$
This inclusion together with the constraint imposed by Thm.~\ref{thm:restriction} do not suffice 
to fully identify
the set $\mathfrak{M}_{A,{\rm SC}}$. We do not know if a singular continuous invariant
measure $\nu$ can be a semiclassical measure. The set of invariant continuous measures 
is dense in $\mathfrak{M}_A$ \cite{sigmund}, so in any case $\mathfrak{M}_{A,{\rm SC}}$ cannot contain
all continuous invariant measures.

%%%%%%%%%%%%%%%%%%%%%%%%%%%%%%%%%%

\subsection{About the Ehrenfest time}
Proposition~\ref{prop:evolution} means that \textbf{any} sequence of
localized states $\{|\varphi_h\rangle\}_{h\to 0}$ evolves towards a sequence of equidistributed states
at the Ehrenfest time $T=\frac{|\log h|}{\lambda}+\O(1)$. To achieve this goal, the prefactor 
$1/\lambda$ defining $T$ is crucial, and cannot be modified
without stronger assumptions on the localization of $|\varphi_h\rangle$.
Indeed, for any $\eps>0$, one can construct a sequence of 
states $|\varphi_h\rangle$ semiclassically localized at the origin, 
such that the evolved states 
$|\psi_h\rangle=\hat A^{(1-\eps)T}|\varphi_h\rangle$ 
are still localized at the same point. Explicitly,
consider the coherent state at the origin $|0\rangle=|0\rangle_h$, and take the sequences 
$$
|\varphi_h\rangle\defi\hat A_h^{-(1-\eps)T/2}|0\rangle, \quad
|\psi_h\rangle=\hat A_h^{(1-\eps)T}|\varphi_h\rangle=\hat A_h^{(1-\eps)T/2}|0\rangle.
$$
At the ``microscopic scale'', the states $|\varphi_h\rangle$ and $|\psi_h\rangle$ are very 
different: the former is stretched along the stable axis, the latter 
along the
unstable axis. However, the ``length'' of both states is of the order $h^{\eps/2}$, 
so this difference of shape is invisible at the measure-theoretic level, and both sequences admit
the semiclassical measure $\delta_0$. 

On the other hand, there exist an infinite sequence of Planck's constants 
$h_k^{-1}\in\N$, $h_k\to 0$ such that starting from the states $\{|\psi_{h_k}\rangle\}$ 
defined above (localized at the origin), the evolved states 
$\{\hat A^{(1+\eps)T}|\psi_{h_k}\rangle\}$ are localized at
the origin as well. These special values of $h$ correspond 
to ``short quantum periods'' of the quantized cat map \cite{BonDB1}. 
Let us remind that for any $h^{-1}\in\N$, 
the quantum cat map $\hat A_h$ is periodic, meaning
that there exists
$P_h\in\N$ and $\theta_h\in[0,2\pi]$ such that $\hat A_h^{P_h}=\e^{\i\theta_h} {\rm Id}_{h}$ \cite{hannay}. 
Besides, there exists an infinite subsequence $h_k\to 0$ for which these periods 
are as short as 
$P_{h_k}=2T_{h_k}+\O(1)$. As a result, one has
$$
\hat A_{h_k}^{(1+\eps)T}|\psi_{h_k}\rangle=
\hat A_{h_k}^{3(1+\eps)T/2}|0\rangle=\e^{\i\theta_{h_{k}}}
\hat A_{h_k}^{3(1+\eps)T/2-P_k}|0\rangle
=\e^{\i\theta_{h_k}}\hat A_{h_k}^{(-1+\eps)T/2+\O(1)}|0\rangle.
$$
The state on the RHS is close to $|\varphi_{h_k}\rangle$ at the microscopic level, and
therefore admits $\delta_0$ for semiclassical measure. In conclusion, the time $(1-\eps)T$ 
would be too short, 
and $(1+\eps)T$ too long to produce transition localized$\to$equidistributed described in
Prop.~\ref{prop:evolution}.

\subsection{More general maps?}
\label{moremaps}
Proposition~\ref{prop:evolution} can be extended to a broad class of
quantized ergodic automorphisms on tori of dimension $2d$ with $d>1$ \cite[Thm.~5.1]{debrievre02}. 
The precise conditions on the automorphism $A$ for the proposition to be
satisfied are the following:
\begin{enumerate}
\item $A$ is ergodic (yet not necessarily hyperbolic), meaning that none of its eigenvalues is
a root of unity.
\item $A$ does not leave invariant any non-trivial proper sublattice of $\Z^{2d}$.
\item Let $\e^{\lambda}$ be the module of the largest eigenvalue(s) of $A$ and $E_{\lambda}$
the direct sum of the generalized eigenspaces corresponding to eigenvalues of modulus $\e^\lambda$. 
Then $A$ restricted to $E_{\lambda}$ must be diagonalizable.
\end{enumerate}
The two last conditions are always satisfied for ergodic automorphisms 
in the case $d=1$, but not necessarily in higher dimension. One easily checks that 
Proposition~\ref{pro:crossed}
(and the Lemma~\ref{lem:nu(stable)} it depends on) also holds for higher-dimensional automorphisms
satifying the above conditions.

As opposed to the proof of \cite[Thm.~1.2]{debrievre02},
our Thm.~\ref{thm:maximal_measure} crucially relies 
on the two dynamical propositions of Section~\ref{s:evolution}, while
the remaining ingredient in the proof of the theorem (namely, the splitting of eigenstates
into two parts) can be straightforwardly transposed to higher dimension. Therefore, 
Theorems~\ref{thm:maximal_measure} and \ref{thm:restriction} also hold in higher dimension for the class of
ergodic automorphisms described above. 
We do not know if the upper bound $1/2$ is sharp in dimension $d>1$; in fact,
the periods of the quantized automorphisms 
are then $>>|\log h|$ (Z.~Rudnick, private communication), 
which makes the construction of \cite{fred-steph02} irrelevant.

\medskip

Back to the 2-dimensional torus, a 
natural extension of the above results would concern the perturbations of
the linear map $A$ of the form $M=\e^{-tX_H}\circ A$, where $X_H$ is the vector field generated by
a Hamiltonian $H\in C^\infty(\t2)$. For $t$ sufficiently small, this map is still Anosov. The challenge
consists in generalizing Propositions~\ref{prop:evolution} and \ref{pro:crossed} to those maps, 
with an appropriate definition
of the Ehrenfest time. The trick \eqref{eq:trick}
used in the linear case to prove these propositions cannot be used for a nonlinear perturbation, 
the problem starting from the poor control of
Egorov's property \eqref{egorov} for times of order $T$.

\medskip

Finally, one may also try to prove a similar property for chaotic flows, {\it e.g.} the geodesic
flow on a compact Riemann surface of negative curvature. In such a setting, some interesting 
results have been recently obtained by R. Schubert, 
pertaining to the long-time evolution of Lagrangian states, which is a first step towards the
proof of Proposition~\ref{prop:evolution} in this setting.

\bigskip

{\bf Acknowledgements:}
We have benefitted from insightful discussions with R.~Schubert,
and Y.~Colin de Verdière, whose remarks also motivated this work. Both authors acknowledge the support
of the Mathematical Sciences Research Institute (Berkeley) where this work was completed.

\appendix

\section{Proof of Lemma \ref{lem:observable-microloc}\label{a:obs-micro}}
We start by showing that for $x$ outside a small disk around $x_i$, the  state
$\hat\vartheta_{i,r(h)}^{\AW}|x\rangle$ 
is asymptotically small (where $|x\rangle$ is a torus coherent state at the point $x$). More precisely, 
there exists a sequence of radii $R(h)\searrow 0$
such that, for any $x_i\in \S$ and
any sequence of points $\{x_h\in \t2\}$ satisfying $x_h\not\in D(x_i,R(h))$, then 
$\norm{\hat\vartheta_{i,r(h)}^{\AW}|x_h\rangle}\leq h^2$ for sufficiently small $h$.

First of all, we recall a couple of estimates on torus coherent states, valid for small enough $h$. 
\begin{itemize} 
\item for any $x\in\t2$, $\norm{|x\rangle}\leq 2$.
\item for any $x,y\in\t2$, one has
$\abs{\langle x|y\rangle}\leq 5\exp\{\pi|x-y|^2/2h\}$, where $\abs{x-y}$ denotes the 
torus distance between the points $x,y$.
\end{itemize}
Now, the operator $\hat\vartheta_{i,r(h)}^{\AW}$ is a combination of projectors 
$|x\rangle\langle x|$ for points
$x$ in the disk $D(x_i,2r(h))$. 
 Therefore, taking $R(h)=2r(h)+\sqrt{2h\abs{\log h}}$ will do the job: any 
$x_h\not\in\nolinebreak D(x_i,R(h))$ and
any $x\in D(x_i,2r(h))$ satisfy $\abs{x-x_h}\geq \sqrt{2h\abs{\log h}}$, which implies 
for $h$ small enough 
$\abs{\langle x|x_h\rangle}\leq 5\exp(-2\pi h\abs{\log h}/2h)\leq h^3$. One finally gets
$$
\norm{\hat\vartheta_{i,r(h)}^{\AW}|x_h\rangle}
\leq\int_{D(x_i,2r(h))}\frac{dx}{h}\;\norm{|x\rangle}\;\abs{\langle x|x_h\rangle}
\leq 2 h^2\;Vol(D(x_i,2r(h)))\leq h^2.
$$
We are now in position to estimate the operator product 
$$
\hat f^{\AW}\,\hat\vartheta_{i,r(h)}^{\AW}
=\int_{\t2}\frac{dy}{h}\;|y\rangle f(y)\langle y|\,\hat\vartheta_{i,r(h)}^{\AW}
$$
by separating the integral into two parts. On the one hand, the integral over
$\t2\setminus D(x_i,R(h))$ is bounded above by $2 h\norm{f}_\infty$ from the above results. 
On the other hand, on $D(x_i,R(h))$ the function $f(y)$ is equal to
the function $f(x_i)+g_h(y)$, where 
$g_h(y)\defi (f(y)-f(x_i))\vartheta_{i,R(h)}(y)$. Since $g_h(y)$  is
uniformly bounded on $\t2$, the same arguments as above yield
$$
\hat f^{\AW}\,\hat\vartheta_{i,r(h)}^{\AW}
=f(x_i)\hat\vartheta_{i,r(h)}^{\AW} +
\hat g_h^{\AW}\,\hat\vartheta_{i,r(h)}^{\AW} +\O(h).
$$ 
The function $g_h$ actually decreases uniformly with $h$
due to the smoothness of $f$: 
$$
\norm{g_h}_\infty\leq \sup_{|y-x_i|\leq 2R(h)}\big(|f(y)-f(x_i)|\big)
\leq 2\,\norm{df}_\infty R(h).
$$ 
This upper bound also
applies to the anti-Wick quantization of $g_h$, so that\\
$\norm{\hat g_h^{\AW}\hat\vartheta_{i,r(h)}^{\AW}}\leq 2\norm{df}_\infty R(h)\htoo 0$.
\qed

\section{Proof of Lemma~\ref{lem:nu(stable)}\label{a:stable}}
We first replace $\S$ by the {\it finite} invariant set it generates, $\S'=\cup_{n\in\Z}A^n(\S)$.
We then want to prove that if $I_0\defi \S'+[a,b]\mathbf{e}^{stable}$ with $0<a<b$,
then $\nu(I_0)=0$ if $\nu$ is an invariant probability measure.
Let $n_0$ be an integer such that $a\e^{\lambda n_0}>b$. Then, the sets
$$
I_j\defi A^{jn_0}(I_0)=\S'+ [a\e^{\lambda jn_0},b\e^{\lambda jn_0}]\mathbf{e}^{stable},\quad
j\in\Z
$$ 
are pairwise disjoint. The invariant measure $\nu$ will satisfy for any $J\geq 0$:
$$
\nu\bigg(\bigcup_{j=-J}^J I_j\bigg)=\sum_{j=-J}^J \nu(I_j)=(2J+1)\nu(I_0).
$$
Since $\nu(\t2)=1$, one must therefore have $\nu(I_0)=0$. 
\qed

This lemma can be easily generalized to the case of the higher-dimensional ergodic toral
automorphisms satisfying the conditions stated in section~\ref{moremaps}. It can also be
extended to any Anosov map $M$ on $\t2$, the straight segments making up $I_0$ being replaced by  
segments of the stable manifolds of a set of periodic points.

%\bibliographystyle{plain}
%\bibliography{/home/faure/articles/articles}

\begin{thebibliography}{99}

\bibitem{BonDB1}F.~Bonechi and S.~De Bi\`evre, 
{\it Exponential mixing and $\ln \hbar$ timescales in quantized hyperbolic maps on the torus}, 
Commun. Math. Phys. {\bf 211}, 659--686 (2000)

\bibitem{debrievre02}
F.~Bonechi and S.~De~Bièvre,
\newblock {\em Controlling strong scarring for quantized ergodic toral automorphisms},
Duke Math. J. (in press)

\bibitem{BDB}A.~Bouzouina and S.~De~Bièvre, {\it Equipartition
of the eigenfunctions of quantized ergodic maps on the torus}, Commun. Math.
Phys. {\bf 178},  83--105 (1996)

\bibitem{schubert} C.-H.~Chang, T.~Kr\"uger, R.~Schubert, 
{\it Quantizations of piecewise affine maps on the torus and their quantum limits}, in preparation

\bibitem{CdV}Y.~Colin de Verdi\`ere, {\em Ergodicit\'e et fonctions
propres du Laplacien}, Commun. Math. Phys. {\bf 102} 497--502, (1985)

\bibitem{DE}M.~Degli~Esposti,
\newblock {\it Quantization of the orientation preserving automorphisms of the torus}, 
Ann. Inst. Henri Poincar\'e {\bf 58}, 323--341 (1993)

\bibitem{fred-steph02}
F.~Faure, S.~Nonnenmacher and S.~De~Bièvre,
\newblock {\em Scarred eigenstates for quantum cat maps of minimal periods,}
Commun. Math. Phys. (in press).

\bibitem{hannay}J.H.~Hannay and M.V.~Berry, {\it Quantization of linear
maps-Fresnel diffraction by a periodic grating}, Physica D {\bf 1}, 267--290
(1980)

\bibitem{heller}
E.J.~Heller, {\em Bound-state eigenfunctions of classically chaotic Hamiltonian systems: 
scars of periodic orbits}, Phys. Rev. Lett. {\bf 53}, 1515--1518 (1984)

\bibitem{K}S.~Koyama, {\em Quantum ergodicity of Eisenstein series for arithmetic 3-manifolds}, 
Commun. Math. Phys. {\bf 215}, 477--486 (2000)

\bibitem{kuru1} P.~Kurlberg and Z.~Rudnick, {\it Hecke theory and
equidistribution for the quantization of linear maps of the torus}, Duke Math. J. {\bf 103}, 47--77 (2000)

\bibitem{linden}E.~Lindenstrauss, {\it Invariant measures and arithmetic quantum unique ergodicity},
preprint (2003)

\bibitem{LS}W.~Luo and P.~Sarnak, {\em Quantum ergodicity of eigenfunctions on $PSL_2(\Z)\setminus H^2$},
Publ. Math. IHES {\bf 81}, 207--237 (1995)

\bibitem{marcus}B.~Marcus, {\em A note on periodic points of toral automorphisms},
Monatsh. Math. {\bf 89}, 121--129 (1980)

\bibitem{MR}J.~Marklof and Z.~Rudnick, {\em Quantum unique ergodicity for parabolic maps},
Geom. Funct. Anal. {\bf 10} 1554--1578 (2000)

\bibitem{PS}Y.~Petridis and P.~Sarnak, {\em Quantum unique ergodicity for $SL_2(\O)\setminus H^3$ and
estimates for $L$-functions}, J. Evol. Equ. {\bf 1}, 277--290 (2001) 

\bibitem{RS}Z.~Rudnick and P.~Sarnak, {\em The behaviour of eigenstates of arithmetic hyperbolic
manifolds}, Commun. Math. Phys. {\bf 161}, 195--231 (1994)

\bibitem{Sc} A.~Schnirelman, {\em Ergodic properties of eigenfunctions},
Usp. Math. Nauk {\bf29}, 181--182  (1974)

\bibitem{sigmund}K.~Sigmund, {\em Generic Properties of Invariant Measures for Axiom A-Diffeormorphisms},
Invent. Math. {\bf 11}, 99--109 (1970)

\bibitem{W} S.~A.~Wolpert, {\em The modulus of continuity for $\Gamma\sb 0(m)\backslash\mathbb{H}$ 
semi-classical limits},  Commun. Math. Phys. {\bf 216}, 313--323 (2001)

\bibitem{Z1} S.~Zelditch, {\em Uniform distribution of the eigenfunctions
on compact hyperbolic surfaces}, Duke Math. J {\bf 55}, 919--941  (1987); 
 {\em Index and dynamics of quantized contact transformations},
Ann. Inst. Fourier {\bf 47}, 673--682 (1996)

\end{thebibliography}

\end{document}